\documentclass[12pt]{article}

\usepackage{amsmath,amsfonts,epsfig,mathrsfs,todonotes,yfonts} 

\usepackage{eufrak}
\usepackage{tikz}
\usetikzlibrary{shapes,arrows}

\usepackage{xcolor}
\usepackage{cite}
\usepackage{stmaryrd}
\usepackage{mdframed}
\usepackage[top=2.5cm, bottom=2.5cm, left=2.75cm, right=2.75cm]{geometry} 
\usepackage{dsfont}
\numberwithin{equation}{section}
\usepackage{hyperref}
 \hypersetup{
     colorlinks=true,
     linkcolor=black,
     filecolor=blue,
     citecolor=blue,      
     urlcolor=cyan,
     }
\usepackage[most]{tcolorbox}

\newcommand{\re}[1] {(\ref{#1})}

\newcommand{\pa}{\partial} 

\newcommand{\ber}{\begin{eqnarray}}
\newcommand{\eer}[1]{\label{#1}\end{eqnarray}}
\newcommand{\eero}{\end{eqnarray}}

\newcommand{\balg}{\begin{align}}
\newcommand{\ealg}{\end{align}}

\newcommand{\beq}{\begin{equation}}
\newcommand{\eeq}{\end{equation}}
\newcommand{\bea}{\begin{eqnarray}}
\newcommand{\eea}{\end{eqnarray}}

\newcommand{\nn}{\nonumber}
\newcommand{\na}{\nabla}

\newcommand{\half}{{\textstyle{\frac12}}}

\def\HollowBox #1#2{{\dimen0=#1 \advance\dimen0 by -#2
       \dimen1=#1 \advance\dimen1 by #2
        \vrule height #1 depth #2 width #2
        \vrule height 0pt depth #2 width #1
        \llap{\vrule height #1 depth -\dimen0 width \dimen1} 
       \hskip -#2
       \vrule height #1 depth #2 width #2}}

\usepackage{fancyhdr} 
\newcommand{\auth}{\large Ulf Lindstr\"om ${}^{a,b}$\footnote{email: ulf.lindstrom@physics.uu.se}
and {\"O}zg{\"u}r Sar{\i}o\u{g}lu ${}^a$\footnote{email: sarioglu@metu.edu.tr}}
\begin{document}
\begin{flushright}
{\small UUITP-43/22}\\
\vskip 1.5 cm
\end{flushright}

\begin{center}
{\Large{\bf Gravitational duality,  Palatini variation and boundary terms: A synopsis}}
\vspace{.75cm}

\auth
\end{center}
\vspace{.5cm}
\vspace{.5cm}
\centerline{${}^a${\it \small Department of Physics, Faculty of Arts and Sciences,}}
\centerline{{\it \small Middle East Technical University, 06800, Ankara, Turkey}}
\vspace{.5cm}
\centerline{${}^b${\it \small Department of Physics and Astronomy, Theoretical Physics, Uppsala University}}
\centerline{{\it \small SE-751 20 Uppsala, Sweden }}

\vspace{1cm}


\centerline{{\bf Abstract}}
\bigskip

\noindent
We consider $f(R)$ gravity and Born-Infeld-Einstein (BIE) gravity in formulations where 
the metric and connection are  treated independently and integrate out the metric to 
find the corresponding models solely in terms of the connection, the archetypical treatment 
being that of Eddington-Schr\"odinger (ES) duality between cosmological Einstein  and Eddington 
theories. For dimensions $D\ne2$, we find that this requires $f(R)$ to have a specific form which 
makes the model Weyl invariant, and that its Eddington reduction is then equivalent to that of 
BIE with certain parameters. For $D=2$ dimensions, where ES duality is not applicable, we find 
that both models are Weyl invariant and equivalent to a first order formulation of the bosonic string. 
We also discuss the form of the boundary terms needed for the variational principle to be well 
defined on manifolds with non-null boundaries. This requires a modification of the 
Gibbons-Hawking-York (GHY) boundary term for gravity. This modification also means that 
the dualities between metric and connection formulations are consistent and include the 
boundary terms.
\vskip .5cm

\vspace{0.5cm}
\small

\renewcommand{\thefootnote}{\arabic{footnote}}
\setcounter{footnote}{0}

\pagebreak
\tableofcontents
\setcounter{page}{2}

\section{Introduction}
There has been a lot of interest in modified theories of gravity, often motivated by the singularities in black hole solutions of General Relativity (GR) or the abundance of dark energy and dark matter. A popular venue for modifying GR has been to treat the connection as an independent variable. This can lead to a dual theory, as in the Eddington-Schr\"odinger (ES) duality, or to a new model, as is the case already in the Palatini approach to GR if the matter coupling leads to a violation of the 
metricity condition\footnote{Historical note: One of us pioneered applying the ``Palatini variational principle'' to certain matter coupled gravity theories \cite{Lindstrom:1975ry,Lindstrom:1976pq}, almost fifty years ago. Not many papers followed on these until the early 2000s, but since then the topic has grown impressively. Now there are many hundred papers, mainly due to applications in astrophysics.}.

In this letter we compare two ways of constructing a purely connection dependent gravity model starting from $f(R)$ theory or from Born-Infeld-Einstein (BIE) theory\footnote{Also known as Eddington-inspired Born-Infeld gravity.}. We find that  certain Weyl invariant  $f(R)$ theories lead to  models that are equivalent to those from BIE theory for $D\ne2$ where they both give actions involving the square root of the determinant of the Ricci tensor. The constructions break down for $D=2$, where they lead to Weyl invariant models still involving both the metric and the connection. These models are then both equivalent to the first order action of the bosonic string. 

For each model we find the appropriate generalizations of the Gibbons-Hawking-York GHY) 
boundary terms that ensure a consistent derivation of the field equations in manifolds with 
boundaries under a variational principle that we specify. These terms  are needed for complete duality.

We give a unified treatment of several models, the novel feature being a description of their 
interrelations as summarized in Fig.1, and some aspects of their variation. In particular, we 
emphasize that the two-dimensional versions are special and related to a first order formulation 
of the bosonic string. In addition we discuss the variational principle in the metric-affine setting 
and introduce new GHY type terms for manifolds with boundary.

The layout of the paper is that in secs. \!\ref{ES}-\ref{EIG} we remind the reader of ES duality, 
$f(R)$ theory and BIE theory, respectively. There we display the purely connection and the dual 
purely metric theories in the cases when they exist separately, as well as the mixed theories. 
Sec. \!\ref{2D} contains the $2D$ string-limit of the theories. In sec. \!\ref{bdy} we discuss 
variational principles and boundary terms. We collect some conclusions in sec. \!\ref{conc}.

\section{Eddington-Schr\"odinger }
\label{ES}
The original form of ES duality starts from the action 
\cite{Schrodinger:2011gqa, Eddington},
\ber
S_{ES}=\half\int d^Dx \sqrt{|g|}\Big(g^{ab}R_{ab}(\Gamma)-\Lambda\Big)
\eer{act2}
with metric and (symmetric) connection as independent variables and $|g|$ the absolute value of the determinant of the metric $g_{ab}$. (The original discussion 
has $D=4$.) Eliminating $\Gamma$ via its field equations determines the connection to be 
the Levi-Civita connection $\Gamma^{(0)}$ via the Palatini variation of the Ricci tensor 
\ber
&&\delta R_{ab}=\na_c\delta \Gamma^c_{~ab}-\na_{(a}\delta\Gamma^d_{~b)d}~.
\eer{pal}
This returns the GR action with a cosmological term when substituted into the action
\ber
S^g_{ES}=\half\int d^Dx \sqrt{|g|}\Big(g^{ab}R_{ab}(\Gamma^{(0)})-\Lambda\Big)~.
\eer{act22}
The boundary contribution generated in the process reads
\ber
\int \na_{[c}\Big(\sqrt{|g|}g^{ab}\delta \Gamma^c_{~a]b}\Big)~.
\eer{ESB}

Varying the metric using
\ber
\delta\sqrt{|g|} =-\half \sqrt{|g|}g_{ab}\delta g^{ab} 
\eer{gvar}
gives the field equations
\ber
R_{ab}-\half g_{ab}\left(g^{cd}R_{cd}-\Lambda\right)=0~.
\eer{ConEq}
When $D\ne 2$ this can be solved to give
\ber
g^{ab}R_{ab}=\frac D {D-2}\Lambda~.
\eer{sec2eq7}
From \re{ConEq} it then follows that 
\ber
g_{ab}\Lambda=(D-2)R_{ab}
\eer{sec2eq8}
and thus
\ber
\sqrt{|g|}=\Big(\frac {D-2}\Lambda\Big)^{\frac D 2}\sqrt{|\det R_{ab}|}~.
\eer{sec2eq9}
The dual theory, with the connection as independent variable is then, from \re{act2},
\ber
S^\Gamma_{ES}=\Big(\frac {D-2} \Lambda \Big)^\frac {D-2} 2\int d^Dx \sqrt{|\det R_{ab}(\Gamma)|}~.
\eer{ES1}

\section{First order \texorpdfstring{$f(R)$}{f(R)} and duality}
The second order $f(R)$ theories in $D$ dimensions are defined by an action
\ber
S^g_f= \int d^Dx \sqrt{|g|} f(R)~,
\eer{act1}
where 
\ber
R(\Gamma^{(0)}) = g^{ab} R_{ab}(\Gamma^{(0)})~.
\eer{sec3eq2}
Here we will consider the action $S_f$, which is \re{act1} but with $R(\Gamma^{(0)})$ 
replaced by $R(\Gamma)$ where $\Gamma$ is a general (symmetric) connection.
Varying the metric and the connection independently as in the Palatini formulation of GR will 
in general result in field equations inequivalent to those from \re{act1}, \cite{Olmo:2011uz}. Here 
we are interested in the possibility of dualising the model to one written entirely in terms of 
the connection. This procedure is related to the ES duality described 
in sec. \!\ref{ES}.

The variation of the general action reads
\ber\nn
\delta S_f&=&\delta\!\!\int\!d^Dx \sqrt{|g|}f(R)\\[1mm]
&=&
\!\!\int\!d^Dx \Big(\sqrt{|g|}~\!\delta g^{ab}\left[-\half g_{ab}f(R)+f'(R)R_{ab}\right]+2g^{ab}f'(R)\sqrt{|g|}\na_{[c}\delta\Gamma^c_{~a]b}\Big)~,
\eer{vact2}
where use is made of the relations \re{pal} and \re{gvar}. Extremising the action will then result in a 
boundary contribution
\ber
2\int\!d^Dx \na_{[c}\Big(g^{ab}f'(R)\sqrt{|g|}\delta\Gamma^c_{~a]b}\Big)
\eer{FRB}
and the field equations
\ber\nn
&&f'(R)R_{ab}-\half g_{ab}f(R)=0 \,, \\[1mm]
&&\na_c\big(\sqrt{|g|}g^{d b}f'(R)\big)-\delta^d_c\na_a\big(\sqrt{|g|} g^{ab}f'(R)\big)=0~.
\eer{feq}
Taking  the determinant and the trace of the first of these results in
\ber
\sqrt{|g|}=\Big|\frac{2f'} f\Big|^{\frac D 2}\sqrt{|\det R_{ab} |}
\eer{detg}
and
\ber
D=\frac{2f'} f R \,,
\eer{ODE}
respectively. The equation \re{ODE} gives a first order ordinary differential equation for $f$ which 
we solve\footnote{Here we are primarily interested in ES duality, but the question of special cases 
is interesting. In \cite{Ferraris:1992dx}, a more detailed analysis of the solutions to \re{feq} is given,
covering also the case e.g. when the first equation of \re{feq} is NOT identically satisfied (see their 
``case 2"). There they do not find any special case where the metric can be eliminated.}
\ber
\frac {df} f=\frac D 2 \frac {dR} R~~~\Rightarrow~~~f(R)=cR^{\frac D 2}
\eer{fRspec}
with $c$ a constant.  Combining this with \re{detg} eliminates the metric completely and 
gives the action 
\ber
S_f=\int\!d^Dx\sqrt{|g|}f(R)~~ \to~~~ S_f^\Gamma=cD^{\frac D 2}\int\!d^Dx \sqrt{|\det R_{ab}(\Gamma )|} \,.
\eer{dual}
The action on the left hand side is invariant under Weyl transformations of the metric
\ber
g_{ab}\to \Omega(X)g_{ab}
\eer{gtog}
in parallel to the construction of Weyl invariant (spinning and super) p-branes
\cite{Lindstrom:1987cv,Lindstrom:1987ps,Lindstrom:1987fr,Lindstrom:1988zm,Lindstrom:1988az}. 
An alternative way to see this is by using \re{detg} and \re{ODE} simultaneously to massage 
the right hand side of $S_f^\Gamma$ in \re{dual} by direct substitution and elementary linear algebra, to arrive at
\ber
c\int\!d^Dx\sqrt{|\det (g_{ab}(g^{cd}R_{cd}))|}~,
\eer{altdual}
which is clearly Weyl invariant by \re{gtog}.

Going back to the field equations \re{feq} and tracing the second equation over $(c,d)$ implies
\ber
\na_c\big(\sqrt{|g|}g^{c b}f'(R)\big)=0~,
\eer{cfe}
in all dimensions $D\ne2$, so that, for the solution \re{fRspec} we have\footnote{At this stage it is common to argue (in $D=4$) that \re{cfe}  leads to a Levi-Civita connection for $q_{ab}:=f'g_{ab}$. Since this still contains the connection, the field equations are then used to replace $R_{ab}$ by the energy momentum tensor. See, e.g., \cite{Olmo:2011uz} or \cite{Saez-ChillonGomez:2020afj}. This route is not open to us since for us $D$ is arbitrary and we have no matter.}
\ber
0=\na_c\big(\sqrt{|g|}g^{d b}f'(R)\big)=\na_c\big(\sqrt{|g|}g^{d b}R^{\frac {D-2}2}\big)
=\na_c\big(\sqrt{|g|}g^{d b}\big)R^{\frac {D-2}2}+\sqrt{|g|}g^{d b}\na_cR^{\frac {D-2}2}.
\eer{IdId}
Tracing on $(d,b)$ and using that
\ber
g_{db}\na_c(\sqrt{|g|}g^{db})=(D-2)\na_c(\sqrt{|g|})~,
\eer{sec3eq14}
we find
\ber\frac{\na_c\sqrt{|g|}}{\sqrt{|g|}}+\frac D 2 R^{-1}\na_cR=0~~~
\Rightarrow~~~\pa_c \ln{\sqrt{|g|}}-\Gamma_c+\frac D 2 {\na_c \ln R}=0~,
\eer{sec3eq15}
where $\Gamma_c$ is the contracted connection. Noting that $\pa_c \ln{\sqrt{|g|}}=\Gamma^{(0)}_c$, the contracted Levi-Civita connection, and defining $R^{-1}\na_cR=:\pa_c\varphi$, we have
\ber
\Gamma_c=\Gamma^{(0)}_c+\frac D 2\pa_c\varphi~.
\eer{Gam3}
Making use of this in \re{IdId} yields
\ber\nn
&&\na_c\big(\sqrt{|g|}g^{d b}\big)+\big(\frac {D-2}2\Big)\sqrt{|g|}g^{d b}R^{-1}\na_cR=0~,\\[1mm]\nn
&&\implies \pa_c(\ln \sqrt{|g|})g^{d b}-\Gamma^{(0)}_cg^{d b}+\na_cg^{d b}-\frac D 2\pa_c\varphi g^{d b}+
\big(\frac {D-2}{2}\Big)g^{d b}\pa_c\varphi=0\\[1mm]
&&\implies \na_cg^{d b}-\pa_c\varphi g^{d b}=0~,
\eer{setup}
where we have used \re{Gam3} in an intermediate step.
The last equation in \re{setup} may be solved in the way the Christoffel symbols are found from the metricity condition to give
\ber
\Gamma^a_{~cd}=:\tilde\Gamma^a_{~cd}=\Gamma^{(0)a}_{~~~~cd}+\half g^{ab}\left(2g_{b(c}\pa_{d)}\varphi - g_{cd }\pa_b\varphi\right)~,
\eer{GAM}
with $\Gamma^{(0)a}_{~~~~cd}$ the Levi-Civita connection\footnote{Formally, we are still 
faced with the same nonlinear problem as discussed in footnote 4. However, now the problem 
is located in the $\varphi$ dependence and we may use Weyl invariance to go to a particular 
gauge where the connection is independent of $\varphi$. 

In $D=2$ the solution contains an independent vector as part of the first order version of the bosonic string in \cite{Lindstrom:1987sg}. There it also corresponds to an additional symmetry of the model, unique to $D=2$.}. This is consistent with \re{Gam3} and shows that the connection $\tilde\Gamma^a_{~cd}$ equals the Levi-Civita connection $\Gamma^{(0)a}_{~~~~cd}$ up to a Weyl transformation 
\ber
g^{d b} \to e^\varphi g^{d b}~.
\eer{sec3eq19}
Since the model is Weyl invariant, we can thus always choose the connection to be Levi-Civita.

In the preceding derivation we only used the equation \re{ODE} from the metric 
field equations to determine the form of $f(R)$. So if we assume that form to begin with, the 
result \re{GAM} implies that the action $S_f^\Gamma$ in \re{dual} is dual to
\ber
S_f^{g}=\int\!d^Dx\sqrt{|g|}R^{\frac D 2}( \Gamma^{(0)} )~.
\eer{sec3son}

\section{Born-Infeld-Einstein Gravity}
\label{EIG}
In this section we relate the previous results to a  gravitational model  based on an analogy 
to the Born-Infeld action. We shall be brief in our presentation, since we recently discovered 
that several of the calculations already exist in the literature (for $D=4$)
\cite{Banados:2010ix,Nascimento:2019qor}.

The action for BIE gravity \cite{Deser:1998rj}, with independent (symmetric) 
connection reads
\ber
S_{BIE}=\frac 1 \epsilon \int d^Dx \Big(\sqrt{|\det(g_{ab} + \epsilon R_{ab}(\Gamma)|}
-\lambda \sqrt{|\det(g_{ab}|}\Big)~,
\eer{act3}
where the Ricci tensor is assumed symmetric and $\epsilon$ is a parameter of dimension (length)${}^2$. Abbreviating
\ber
h_{ab}:= g_{ab} + \epsilon R_{ab}~,~~~h^{ab}h_{bc}=\delta^a_c~,
\eer{metric2}
and extremising $S_{BIE}$ by varying the metric and the connection using \re{pal} and \re{gvar},  
yields a boundary contribution
\ber
\int d^Dx\na_{[c}(\sqrt{|h|}h^{ab}\delta\Gamma^c_{~a]b})
\eer{EiBIB}
along with the field equations
\ber
\sqrt{|h|}h^{ab}-\lambda \sqrt{|g|} g^{ab}=0
\eer{f1}
and
\ber
\na_{c}(\sqrt{|h|}h^{d b})-\delta^d_c\na_{a}(\sqrt{|h|}h^{ab})=0~.
\eer{f2}
When $D\ne 2$ we can solve \re{f1} to give
\ber
\sqrt{\frac{|h|}{|g|} }=\lambda ^\frac{D}{D-2}
\eer{sec4eq6}
which substituted back yields
\ber
g_{ab}=\lambda^\frac{-2}{D-2}h_{ab}=\lambda^\frac{-2}{D-2}(g_{ab} + \epsilon R_{ab})~,
\eer{geqn}
so that
\ber
g_{ab}=\frac {\epsilon R_{ab}(\Gamma)}
{(\lambda^\frac2{D-2}-1)}~,
\eer{gtoR}
thus eliminating the metric in favour of the connection. Inserting this in \re{act3}, we find an action 
with the connection as the only variable
\ber
S^\Gamma_{BIE}=\frac 1 \epsilon \int d^Dx\big(\sqrt{|h|}-\lambda \sqrt{|g|}\big)=\epsilon^{{\frac D 2 }-1}(\lambda^\frac2{D-2}-1)^{\frac {2-D}2} \int d^Dx\sqrt{|\det R_{ab}(\Gamma)|}~.
\eer{EDD} 
For 
\ber
\lambda =\Big[\frac {\epsilon\Lambda +D-2}{D-2}\Big]^{\frac{D-2}2}~,
\eer{Lrel}
this agrees with the ES result in \re{ES1}\footnote{For $D=4$ \re{Lrel} implies 
$2(\lambda-1)= \epsilon\Lambda$.}.
The two models are thus equivalent as starting points for deriving the $R(\Gamma)$ theory, 
although only one of them is Weyl invariant. 

Finding a metric action dual to \re{EDD} by solving \re{f2} meets with several obstacles. 
The solution to \re{f2} is formally
\ber
\Gamma=\Gamma^{(0)}(h_{ab})~,
\eer{Gamh}
i.e. the Levi-Civita connection for $h_{ab}$. But this is incomplete since $h_{ab}$ still contains 
$R_{ab}(\Gamma)$. So it is rather the full metric and connection theory with field equations 
\re{f1} {\em and} \re{f2} one has to consider.

Using \re{gtoR} we write
\ber
h_{ab} := g_{ab} + \epsilon R_{ab}=\lambda^\frac{2}{D-2}g_{ab} ~.
\eer{sec4eq12}
Since this is a constant Weyl transformation, plugging it into \re{Gamh} gives
\ber
\Gamma^a_{~bc}=\Gamma^{(0)}(h_{ab})=\Gamma^{(0)}(g_{ab})~,
\eer{Chris}
in perfect agreement with the right hand side of \re{f1}. The on-shell theory thus found 
satisfies \re{gtoR} with connection \re{Chris} is
\ber
{(\lambda^\frac2{D-2}-1)}g_{ab}= {\epsilon R_{ab}\big(\Gamma^{(0)}(g_{ab})\big)}~,
\eer{gtoRR}
which is equivalent to Einstein's vacuum field equations with a cosmological constant
\ber
\half(2-D)\big(\lambda^\frac 2{D-2}-1\big)~.
\eer{sec4eq15}

This confirms the $D=4$ discussion of the matter coupled system in \cite{Banados:2010ix,Nascimento:2019qor} where it  is shown that the vacuum solutions 
are the Einstein solutions.

We saw in Sec.\ref{ES}  that in $D=4$ the $\Gamma$-action \re{dual} is dual to GR with a 
cosmological constant $-\half\Lambda =  -\frac  1{16 c}$. For general $D$, it corresponds 
to BIE with cosmological term 
\[ \lambda =\Big[\frac {\epsilon\Lambda +D-2}{D-2}\Big]^{\frac{D-2}2} \] 
We thus have an interesting chain of equivalences between ``metric'', ``affine'' and ``metric/affine'' models. See the figure below for the interrelations.\\
\bigskip

\tikzstyle{decision} = [diamond, draw, fill=blue!20, 
    text width=4.5em, text badly centered, node distance=3cm, inner sep=0pt]
\tikzstyle{block} = [rectangle, draw, fill=blue!20, 
    text width=5em, text centered, rounded corners, minimum height=4em]
\tikzstyle{line} = [draw, -latex']
\tikzstyle{cloud}= [draw, ellipse,fill=red!20, node distance=4cm,
    minimum height=3em]
\noindent
\bigskip

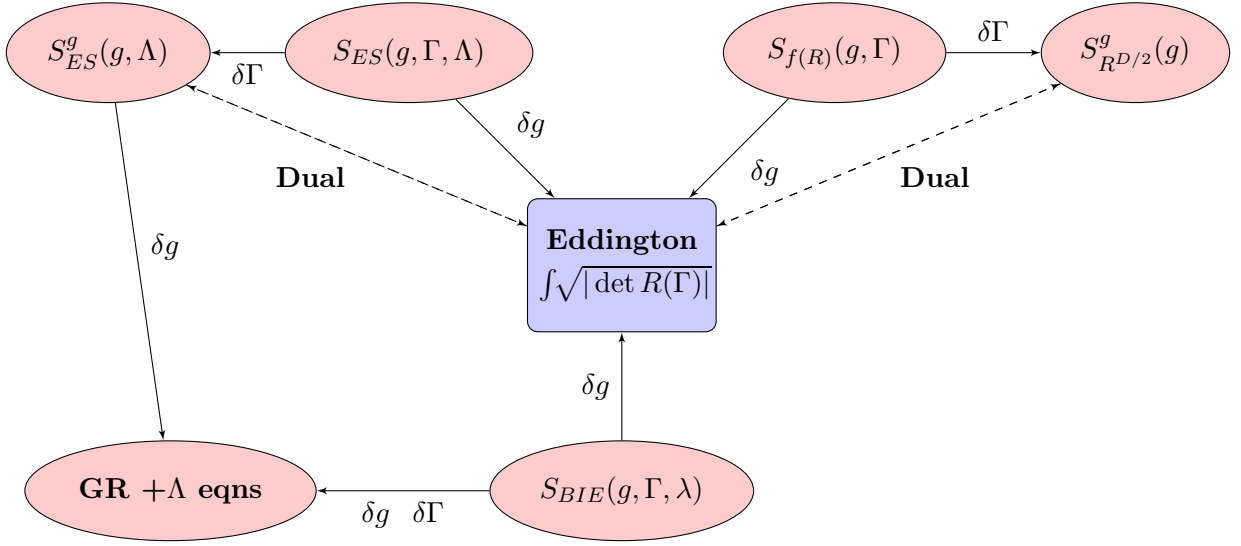
{\bf Fig.1.  A schematic picture of the interrelation between the models discussed:
\bigskip

\noindent
\begin{tikzpicture}[node distance = 2cm, auto]
    \node [block] (init) {Eddington 
    $\int\!\!\sqrt{|\det R(\Gamma)|}$};
    \node [cloud, above left  of=init] (expert) { $S_{ES}(g,\Gamma,\Lambda)$};
    \node [cloud, left  of=expert] (expert2) {$S^g_{ES}(g,\Lambda)$};
    \node [cloud, above right of=init] (system) {$S_{f(R)}(g,\Gamma)$};
    \node [cloud, right of=system] (system2) { $S^g_{R^{D /2}}(g)$};
    \node [cloud, below of=init, node distance=3cm] (evaluate) {$S_{BIE}(g,\Gamma,\lambda)$};
   \node [cloud, left of=evaluate, node distance=6cm] (update) {GR +$\Lambda$ eqns};
    \path [line] (expert)--node {$\delta\Gamma$} (expert2);
    \path [line] (evaluate) -- node {$\delta g~~\delta\Gamma$} (update);
    \path [line] (evaluate) -- node {$\delta g$} (init);
    \path [line] (system) -- node {$\delta\Gamma$} (system2);
    \path [line] (expert) -- node {$\delta g$} (init);
    \path [line] (system)--node {$\delta g$} (init) ;
    \path [line,dashed] (init) -- node {Dual}(expert2);
     \path [line,dashed] (expert2) -- (init);
    \path [line,dashed] (init) --  (system2);
     \path [line,dashed] (system2) -- node  {Dual} (init);
    \path [line] (expert2) -- node {$\delta g$} (update);
\end{tikzpicture}
}

\section{The \texorpdfstring{$2D$}{2D} string limits}\label{2D}
Clearly the results in  the previous sections do not hold when $D=2$.
When $D=2$, the discussion of the model \re{act3} is as follows: Taking determinants in equation 
\re{f1} leads to 
 \ber
 \lambda^2=1~~~\Rightarrow~~~\lambda=\pm 1~,
 \eer{seceq1}
 so that equation \re{geqn} gets replaced by
 \ber
 \pm g^{ac}h_{cb}=\pm (\delta^a_b+\epsilon g^{ac}R_{cb})
 \eer{bosact}
 leading to
 \ber
 \sqrt{|h|}-\lambda \sqrt{|g|}=\frac {\epsilon\lambda} 2\sqrt{|g|}g^{ab}R_{ab}(\Gamma)~.
 \eer{conf}
 
 Alternatively we can see this directly in the action  \re{act3} using that in $D=2$
 \ber
 R_{ab}-\half g_{ab}R=0~
 \eer{sec5eq4}
 for a symmetric Ricci tensor and connection. Then,
 \ber
S_{BIE}=\frac 1 \epsilon \int d^2x \Big(\sqrt{|\det (g_{ab} +  \epsilon R_{ab}(\Gamma)|}
-\lambda \sqrt{|g|}\Big)= \frac 1 \epsilon \int d^2x\sqrt{|g|}\Big(1+\half \epsilon R-\lambda\Big)~,
\eer{act333}
which is the ES action \re{act2} with $\Lambda =\frac 2 \epsilon(\lambda - 1)$.
 
However the action \re{act2} is again equivalent  to the right hand side of \re{conf} since tracing 
the field equations \re{ConEq} now results in
\ber
g^{ab}R_{ab}-\left(g^{cd}R_{cd}-\Lambda\right)=0~~~~\Rightarrow \Lambda =0~.
\eer{ConEq2}
Plugging this back into the action \re{act2} (for $D=2$) again returns the mixed variable 
Lagrangian density \re{conf}\footnote{With $\lambda\epsilon=2$.}.
Thus, unlike the case $D\ne 2$, both models depend on both metric and connection as independent 
variables. Indeed, the actions \re{act2} and \re{act3} in two dimensions reproduces the conformally 
invariant first order action for the bosonic string, introduced in \cite{Lindstrom:1987sg}. At first sight 
this action may look strange, since we are taught in string theory to ignore the $2D$ curvature term 
because it is a total derivative. This is indeed true for the curvature for the spin connection, or 
equivalently when the connection is Levi-Civita. It is not true, however, for the curvature of an arbitrary 
affine connection.
  
Briefly, \re{conf} is invariant, up to a boundary contribution\footnote{The transformation 
\re{Gsym} of $\Gamma$ has the form of a conformal transformation and the usual formulae 
for the variation of the Ricci tensor may be used to evaluate it and find  
\( \delta \big( \sqrt{|g|} g^{ab} R_{ab} \big) = -2 \nabla_a (\sqrt{|g|} V^a) \).}, under Weyl rescaling of the 
metric and transformations of the affine connection that read
\ber
\Gamma^a_{~bc}\to\Gamma^a_{~bc}+2V_{(b}\delta^a_{c)}-g_{cb}V^a~.
\eer{Gsym}
The $\Gamma$ field equations lead to 
\ber
\na_a(\sqrt{|g|}g^{bc})=0
\eer{sec5eq8}
as usual, but unlike when $D\ne2$, the solution is not the Levi-Civita connection $\Gamma^{(0)}$, but 
\ber
\Gamma^a_{~bc}=\Gamma^{(0)a}{}_{bc}+2U_{(b}\delta^a_{c)}-g_{cb}U^a~,
\eer{sec5eq9}
with $U^a$ the most general parameter of the transformation \re{Gsym}. It can then be shown that 
there is a gauge choice where the model becomes the usual Nambu-Goto string.

\section{Boundary actions}
\label{bdy}
In this section we consider what boundary terms need to be added to the actions to cancel the 
respective contributions  \re{ESB},\re{FRB} and \re{EiBIB} and make the variational problem well 
defined in the presence of a non-null boundary.

\subsection{Variational principle}
We are concerned with both the variation of the metric $\delta g$ and the independent
connection $\delta\Gamma$. When the full set of field equations hold, $\Gamma$ gets 
expressed in terms of derivatives of the metric. In this sense the system is reminiscent of 
a Hamiltonian system $H(q,p)$ with the metric corresponding to the coordinates $q$ 
and $\Gamma$ to the momenta $p$. The field equation that gives $\Gamma=\Gamma(g,\pa g)$ 
then roughly corresponds to $\pa H/\pa p=\dot q$ which may be solved for $p=p(q,\dot q)$.

As in the variational principle for $H$, where the derivation of the canonical equations requires 
$\delta q=0$ 
on the boundary of the integration volume while no conditions are required for $\delta p$, we 
shall require $\delta g=0$ on the boundary but leave $\delta\Gamma$ free. This choice can 
also be motivated by the fact that on-shell the connection becomes the Levi-Civita connection 
for the metric and its variation does not vanish on the boundary as well as by the wish to find 
a dual action reproducing the boundary terms for both the dual models.

Apart from the extra variation of $\Gamma$, our derivation follows that  
of the GHY boundary term \cite{York:1972sj,Gibbons:1976ue} described in \cite{Feng:2017ygy}.

The boundary is defined by
\ber
x^a = x^a(y^i) \,, \qquad a = 1, \dots, D \,, \qquad i = 1, \dots, D-1 \,, 
\eer{sec6eq1}
which leads to the tangential vectors to the boundary
\ber
e^a_i=\Big(\frac{\partial x^a}{\partial y^i}\Big)_{\pa{}{\cal M}}
\eer{sec6eq2}
and a normal $n_a$
\ber
e^a_in_a=0~.
\eer{sec6eq3}
The induced  metric is
\ber
\gamma_{ij}=g_{ab}e^a_ie^b_j~.~~~
\eer{sec6eq4}
It is related to the full metric as
\ber
g_{ab}=\gamma_{ab}-n_an_b~,~~~g^{ab}=\gamma^{ab}+n^an^b~.
\eer{sec6eq5}
Gauss' divergence theorem reads
\ber
\int_{\cal M}d^Dx\pa_{a}(\sqrt{|g|}{V^a})= \int_{\pa{}\cal M} d^{D-1}y\sqrt{\gamma}~\!n_a{V^a}~.
\eer{sec6eq6}
In the variational principle we shall keep the normal to the boundary ${\pa{}{\cal M}}$, as well 
as its partial derivatives, constant
\ber
\delta n_a=0~,~~~\delta n^a =0~,~~\delta n_a,_b=0~,~~~\delta n^a,_b =0~.
\eer{sec6eq7}
See sec.4 of \cite{Feng:2017ygy} for details.
We also require that the
metric is held constant when confined to the boundary
\ber
\delta g_{ab}{|_{\pa{}\cal M}}=0~,~~\delta g^{ab}{|_{\pa{}\cal M}}=0~.
\eer{delg}
This means that the induced metric $\gamma_{ij}$ is fixed during the variation. It also implies that,
although $\delta\pa_{c}g_{ab}$ does not vanish on ${\pa{}\cal M}$, the tangential derivatives must also vanish:
\ber
\delta g_{ab ,c }e_{i}^{c}=0~.
\eer{sec6eq9}
It follows that 
\ber
\gamma^{ij}e_{i}^{a }e_{j}^{b}\delta g_{c b ,a}=0~.
\eer{sec6eq10}

\subsubsection{Gibbons-Hawking-York}
The GHY boundary action for GR is\footnote{To make the action finite it is customary to also include $K_0$, the extrinsic curvature of the boundary embedded in flat space. In what follows we omit such terms.}
\ber
\half \int_{\pa{}\cal M}d^{D-1}y \sqrt{\gamma} K~,
\eer{sec6eq11}
where 
\ber
K=g^{ab}K_{ab}=g^{ab}\na_{a}n_b
\eer{TR}
is the trace of the extrinsic curvature (second fundamental form) $K_{ab}$.
There are various definitions of $K_{ab}$. One involving the Lie derivative reads
\ber
K_{ab}=\half{\cal L}_n\gamma_{ab}=\half n^c\na_c\gamma_{ab}+\gamma_{c(a}\na_{b)}n^c~.
\eer{sec6eq13}
This has a trace which is only equivalent to that used in \re{TR} when the connection is the Levi-Civita one. 
\subsubsection{Eddington-Schr\"odinger}
For a general $\Gamma$ the boundary term will have  to be modified. To this end we note that
\ber
\half \int_{\pa{}\cal M}d^{D-1}y \sqrt{\gamma} (g^{ab}\na_{a}n_b+\na_{a}n^a)=: \half\int_{\pa{}\cal M}d^{D-1}y\sqrt{\gamma} ~\hat K
\eer{bdyact1}
reduces to the GHY boundary action when the connection is metric. Further varying the connection
\ber
\half\delta\int_{\pa{}\cal M}d^{D-1}y \sqrt{\gamma} ~\hat K
=\half\int_{\pa{}\cal M}d^{D-1}y \sqrt{\gamma} ~ \delta\hat K~,
\eer{sec6eq15}
we find 
\ber
\delta\hat K=\delta\big(g^{ab}K_{ab}+\na_an^b\big)=-2\delta \Gamma^{[ca]}_{~~~a}n_c
\eer{sec6eq16}
so that \re{bdyact1} precisely cancels the boundary contribution \re{ESB} that arises in the 
$\Gamma $ variation of the ES bulk action \re{act2}.

\subsubsection{\texorpdfstring{$f(R)$}{f(R)}}
Similarly, we can cancel the boundary contribution \re{FRB} that arises in the 
$\Gamma $ variation of the $f(R)$ bulk action \re{act1} by the varying the following boundary action
\ber
\half\int_{\pa{}\cal M}d^{D-1}y \sqrt{\gamma} f'(R)~\hat K~,
\eer{bdyact2}
provided that we also set $\delta R=0$ on the boundary. 
 When the connection is Levi-Civita, this reduces to the boundary term in \cite{Guarnizo:2010xr}. The vanishing of the variation of $R$ is required also there and in \cite{Dyer:2008hb}. We refer the reader 
 to these references for a discussion. In both references the connection is the Levi-Civita connection, 
 so the implications of  $\delta R=0$  merits to be studied in more detail.
 
In the case of duality, i.e. when we can eliminate the connection as an independent variable, the 
connection is the Levi-Civita connection and the counterterm becomes that of the purely metric theory as mentioned above.  Similarly, on the dual side, when we eliminate the metric using 
\re{feq}--\re{fRspec}, the variation of the boundary action \re{bdyact2} becomes
\ber
 \half c D^{D/2}
\int \nabla_{[c} \left( \sqrt{\det R_{ab}|} \, R^{de} \delta \Gamma^{c}{}_{d]e} \right) \,, \eer{new618}
with $R^{de}$ the inverse of the affine Ricci tensor, exactly cancelling the variation of the 
Eddington action \re{dual}. Hence for the dual theories, the counterterm action \re{bdyact2} 
produces the correct counterterms, purely metric and purely affine, respectively, for the
two formulations.

\subsubsection{Born-Infeld-Einstein gravity}
Formally, we can cancel the boundary contribution \re{EiBIB} that arises in the 
$\Gamma $ variation of the BIE bulk action \re{act3} by varying the following boundary action
\ber
\half\int_{\pa{}\cal M}d^{D-1}y \sqrt{\gamma^h} ~\hat K^h~,
\eer{bdyact3}
i.e., the same boundary action as \re{bdyact1} but
with the induced metric $\gamma^h$ and $\hat K^h$  now defined with respect to the metric \re{metric2} so that
\ber
\gamma^h{}_{ij}= h_{ab}e^a_ie^b_j= (g_{ab} + \epsilon R_{ab})e^a_ie^b_j~,~~~\hat K^h=(h^{ab}\na_{a}n_b+\na_{a}n^a)~,~~~n^a=h^{ab}n_b~.
\eer{indmetric2}
The conditions \re{delg} become
\ber
\delta h_{ab}{|_{\pa{}\cal M}}=\delta (g_{ab} + \epsilon R_{ab}){|_{\pa{}\cal M}}=0~,~~~\delta h^{ab}{|_{\pa{}\cal M}}=0~.
\eer{dellh}
Since \re{delg} also holds, this means that we must set
\ber
\delta R_{ab}{|_{\pa{}\cal M}}=0~,
\eer{delRicci}
in analogy to the $f(R)$ case in \re{bdyact2} described above. This is consistent  with the equivalence of BIE theory with $f(R)$ for  the special case \re{fRspec}.

\subsubsection{The bosonic string}
Finally we discuss the boundary terms for the $D=2$ first order bosonic string. This differs from \re{bdyact1} by a term proportional to $U^a$ 
\ber
\half\int_{\cal \pa{}M}dy\sqrt{\gamma}\big(\hat K+2U^an_a\big)~.
\eer{Kheadact}
Here $\hat K$ takes care of the boundary contributions from the bulk variation of $R$ as before, while the action is still invariant under conformal symmetry of $\Gamma$
 \ber
 \Gamma^{a}_{~~bc} \to \Gamma^{a}_{~~bc} +2V_{(b}\delta^a_{c)}-g_{bc}V^a~.
  \eer{confsy}
  This is because
  \ber
  K\to K-g^{ab}\big(2V_{(a}\delta^c_{b)}-g_{ab}V^c
 \big)n_c=K
  \eer{KtoKK}
  while
  \ber
  U_a  \to U_a+2V_{(a}\delta^b_{b)}-g_{ab}V^b=U_a+2V_a
  \eer{2ndBdy}
 and from the $R$ term in the bulk
  \ber
  \half \sqrt{|g|} R\to  \half \sqrt{|g|} R-\half\pa_{a}\big(4\sqrt{|g|}  V^a\big)
  \eer{kokgRtoR}
  which gives a boundary contribution
  \ber
  -\sqrt{\gamma}V^a n_a
  \eer{Vana}
that will cancel the term from \re{2ndBdy}.  So the $U$ boundary term ensures the invariance under \re{confsy} and by the same token takes care of a general variation of $U$ to make the variational principle well defined.

\section{Conclusions}
\label{conc}
We have studied first order models for $f(R)$ theories and BIE theory with metric and connection as independent variables,  and compared them to the famous ES duality. We have shown that, when the metric is eliminated, certain Weyl invariant $f(R)$ models result in dual models that are equivalent to those of ES. Similarly, for BIE  we showed that eliminating the metric again leads to the ES result. In the  form where the connection is the only variable, BIE is thus equivalent to Weyl invariant  $f(R)$ in its dual form.

These constructions hold when the space-time dimension $D>2$ but break down for
$D=2$. The models are then still equivalent but now equal to the first order bosonic string.

We have discussed boundary terms for all these theories in general (no specific form of $f(R)$). After describing our variational principle, we showed that the necessary boundary terms carry over from the connection form to the metric form under duality.

Open problems are to include torsion in the discussion as done for e.g. metric modified gravity in \cite{Olmo:2011uz}, to understand if and how the Weyl invariance of our $f(R)$ model can be related to symmetries of the BIE ation.
\bigskip

\noindent{\bf Acknowledgments}\\
The research of U.L. is supported in part by the 2236 Co-Funded 
Scheme2 (CoCirculation2) of T\"UB{\.I}TAK (Project No:120C067)\footnote{\tiny However 
the entire responsibility for the publication is ours. The financial support received from 
T\"UB{\.I}TAK does not mean that the content of the publication is approved in a scientific 
sense by T\"UB{\.I}TAK.}.

\end{document}